\documentclass[final,11pt, 3p]{elsarticle}
\usepackage{graphics, color, hyperref}
\usepackage{subfig, ulem}
\usepackage{lineno}
\usepackage{multirow}
\newcommand{\1}[1]{\, \mathrm{#1}} % unit
\newcommand{\n}[1]{\mathrm{#1}} % normal (roman) text in math mode

\newcommand{\Qz}{\mbox{CUORE-0}}

\begin{document}
\journal{Advances in High Energy Physics}

\title{Searching for neutrinoless double-beta decay of $^{130}$Te with CUORE}

\author[USC,LNGS]{D.~R.~Artusa}
\author[USC]{F.~T.~Avignone~III}
\author[INFNLegnaro]{O.~Azzolini}
\author[LNGS]{M.~Balata}
\author[BerkeleyPhys,LBNLNucSci,LNGS]{T.~I.~Banks}
\author[INFNBologna]{G.~Bari}
\author[LBNLMatSci]{J.~Beeman}
\author[Roma,INFNRoma]{F.~Bellini}
\author[INFNGenova]{A.~Bersani}
\author[Milano,INFNMiB]{M.~Biassoni}
\author[Milano,INFNMiB]{C.~Brofferio}
\author[LNGS]{C.~Bucci}
\author[Shanghai]{X.~Z.~Cai}
\author[INFNLegnaro]{A.~Camacho}
\author[LNGS]{L.~Canonica}
\author[Shanghai]{X.~G.~Cao}
\author[Milano,INFNMiB]{S.~Capelli}
\author[INFNMiB]{L.~Carbone}
\author[Roma,INFNRoma]{L.~Cardani}
\author[Milano,INFNMiB]{M.~Carrettoni}
\author[LNGS]{N.~Casali}
\author[Milano,INFNMiB]{D.~Chiesa}
\author[USC]{N.~Chott}
\author[Milano,INFNMiB]{M.~Clemenza}
\author[Genova]{S.~Copello}
\author[Roma,INFNRoma]{C.~Cosmelli}
\author[INFNMiB]{O.~Cremonesi\corref{cor_Cremonesi}}
% ...

\address[USC]{Department of Physics and Astronomy, University of South Carolina, Columbia, SC 29208 - USA}
\address[LNGS]{INFN - Laboratori Nazionali del Gran Sasso, Assergi (L'Aquila) I-67010 - Italy}
\address[INFNLegnaro]{INFN - Laboratori Nazionali di Legnaro, Legnaro (Padova) I-35020 - Italy}
\address[BerkeleyPhys]{Department of Physics, University of California, Berkeley, CA 94720 - USA}
\address[LBNLNucSci]{Nuclear Science Division, Lawrence Berkeley National Laboratory, Berkeley, CA 94720 - USA}
\address[INFNBologna]{INFN - Sezione di Bologna, Bologna I-40127 - Italy}
\address[LBNLMatSci]{Materials Science Division, Lawrence Berkeley National Laboratory, Berkeley, CA 94720 - USA}
\address[Roma]{Dipartimento di Fisica, Sapienza Universit\`a di Roma, Roma I-00185 - Italy}
\address[INFNRoma]{INFN - Sezione di Roma, Roma I-00185 - Italy}
\address[INFNGenova]{INFN - Sezione di Genova, Genova I-16146 - Italy}
\address[Milano]{Dipartimento di Fisica, Universit\`a di Milano-Bicocca, Milano I-20126 - Italy}
\address[INFNMiB]{INFN - Sezione di Milano Bicocca, Milano I-20126 - Italy}
\address[Shanghai]{Shanghai Institute of Applied Physics, Chinese Academy of Sciences, Shanghai 201800 - China}
\address[Genova]{Dipartimento di Fisica, Universit\`a di Genova, Genova I-16146 - Italy}
% ...

\date{\today}

\begin{abstract}
Neutrinoless double-beta ($0\nu\beta\beta$)~decay is a hypothesized lepton-number-violating process that offers the only known means of asserting the possible Majorana nature of neutrino mass. 
The Cryogenic Underground Observatory for Rare Events~(CUORE) is an upcoming experiment designed to search for $0\nu\beta\beta$~decay of $^{130}$Te using an array of 988 TeO$_2$ crystal bolometers operated at 10~mK. The detector will contain 206~kg of $^{130}$Te and have an average energy resolution of 5~keV; the projected $0\nu\beta\beta$~decay half-life sensitivity after five years of live time is $1.6\times 10^{26}$~y at $1\sigma$ ($9.5\times10^{25}$~y at the 90\% confidence level), which corresponds to an upper limit on the effective Majorana mass in the range 40--100~meV (50--130~meV). 
In this paper we review the experimental techniques used in CUORE as well as its current status and anticipated physics reach.
\end{abstract}

\begin{keyword}

% Keywords
Neutrinoless double-beta decay \sep Majorana \sep Neutrino mass \sep Bolometers \sep TeO$_{2}$

% PACS codes
\PACS 
23.40.-s \sep %Beta decay 
14.60.Pq \sep %Neutrino mass and mixing
14.60St \sep %Non-standard-model neutrinos, right-handed neutrinos, etc.
07.20.Mc \sep %Cryogenics, low-temperature techniques
07.57.Kp %Bolometers 

%% MSC codes 
% \MSC code \sep code
% or \MSC[2008] code \sep code (2000 is the default)

\end{keyword}

\maketitle

\tableofcontents

% !TEX root = CUORE_status_review_2014.tex
\section{Introduction}\label{sec:intro}

The discovery of neutrino oscillations revealed that neutrinos are massive particles and thereby provided the first evidence of physics beyond the Standard Model~(cf.~\cite{PDG2012,bellini_neutrino_2013}). This development resolved some longstanding mysteries but it also raised anew questions about the fundamental nature of neutrinos, namely: What is the absolute mass scale of the neutrino? What is the hierarchy of its different mass states? Is the neutrino its own antiparticle?

Neutrinoless double-beta ($0\nu\beta\beta$)~decay has attracted a great deal of attention in recent years because of its unique potential to provide insight into the above issues. This lepton-number-violating process, 
%in which two simultaneous beta decays occur in the same nucleus without emission of any final-state neutrinos,
\[
%0\nu\beta\beta:
\quad (Z,A) \rightarrow (Z\!+\!2, A) + 2e^- ~,
\]
can occur only if neutrinos are massive Majorana particles---i.e., if they are their own antiparticles, a possibility first suggested by Ettore Majorana in 1937~\cite{Majorana}. 
Indeed, $0\nu\beta\beta$~decay offers the only feasible means of investigating this question at present~(cf.~\cite{Avignone_NDBD_2008,bilenky_NDBD_2012}). 
If $0\nu\beta\beta$~decay occurs it is extremely rare, with a half life greater than $10^{25}$~years.
Observation of the process would unambiguously establish that neutrinos have Majorana mass and reveal them to be different from the other known fermions, which are Dirac particles. The scenario in which neutrinos are Majorana particles is widely viewed as more ``natural'' from a theoretical standpoint, and if true it would have profound implications for our understanding of how neutrinos acquire mass and possibly how the universe's matter-antimatter asymmetry arose~\cite{PDG2012}.
Experimental searches for $0\nu\beta\beta$~decay also have the potential to provide information about the neutrino mass hierarchy and absolute mass scale, depending on whether $0\nu\beta\beta$~decay is observed and at what sensitivity.

%(Two-neutrino double-beta ($2\nu\beta\beta$)~decay is allowed within the Standard Model and has been observed in many isotopes.)
%\[
%2\nu\beta\beta: 
%\quad (Z,A) \rightarrow (Z\!+\!2, A) + 2e^- + 2\bar{\nu}_e ~.
%\]
%If neutrinos are Majorana particles, one can imagine the $\bar{\nu}_e$ emitted from one of the beta-decay vertices undergoing a helicity flip and being absorbed at the other vertex as a $\nu_e$. In this case there are no final-state neutrinos, thus yielding $0\nu\beta\beta$~decay,

To date there has been only one claim of observation of $0\nu\beta\beta$~decay~\cite{Klapdor:2004wj, Klapdor:2006ff}, which stands in tension with more recent null results from other experiments~\cite{Auger:2012ar,Gando:2012zm,Agostini:2013mzu}.
There are currently at least ten experiments aiming to search for $0\nu\beta\beta$~decay in almost as many candidate isotopes. 
Among them is the Cryogenic Underground Observatory for Rare Events~(CUORE)~\cite{Arnaboldi:2002du,Ardito:2005ar}, which will search for $0\nu\beta\beta$~decay of $^{130}$Te by operating TeO$_2$ crystals as cryogenic bolometers at the underground Laboratori Nazionali del Gran Sasso (LNGS), Italy. 
The current lower limit on the half-life for $0\nu\beta\beta$~decay of $^{130}$Te was established by a predecessor experiment, Cuoricino, at $2.8\times 10^{24}$~y (90\% C.L.)~\cite{Cuoricino_NDBD_2011}. 
CUORE aims to improve on this sensitivity by more than a factor of 30 by operating a larger, cleaner, better-shielded detector with enhanced energy resolution inside a new custom-built cryostat.

In this paper we review the design, status, and physics outlook for CUORE. Section~\ref{sec:bolometers} describes the experimental techniques used, with a special focus on energy resolution (Sec.~\ref{sec:fwhm}) and sources of background (Sec.~\ref{sec:bkg}). Section~\ref{sec:q0} describes the construction and operation of CUORE-0, a prototype detector which is now taking data. In Section~\ref{sec:status} we discuss the status of CUORE, including the ongoing assembly of its detectors and commissioning of its cryogenics, and in Section~\ref{sec:conclusion} we examine the potential physics reach of the experiment.

% !TEX root = CUORE_status_review_2014.tex
\section{Bolometric technique}
\label{sec:bolometers}

\begin{figure}[t]
\begin{centering}
 \begin{tabular}{l c}
 \subfloat[]{\label{fig:tower}\includegraphics[width=0.55\columnwidth]{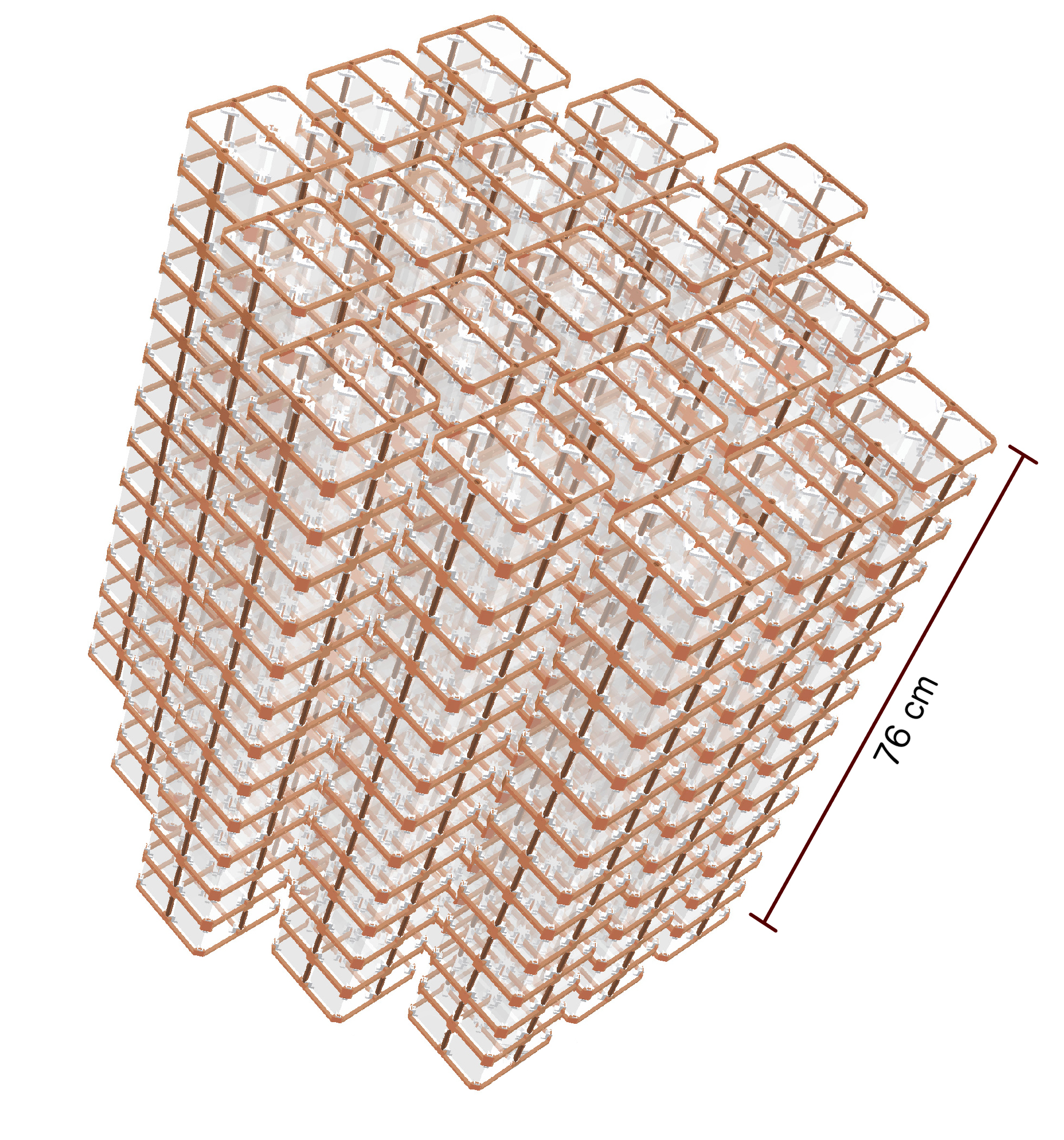}} &
   \hspace{0.02\columnwidth}
 \multirow{-28}[6]{*}{\subfloat[]{\label{fig:2by2}\includegraphics[width=0.35\columnwidth]{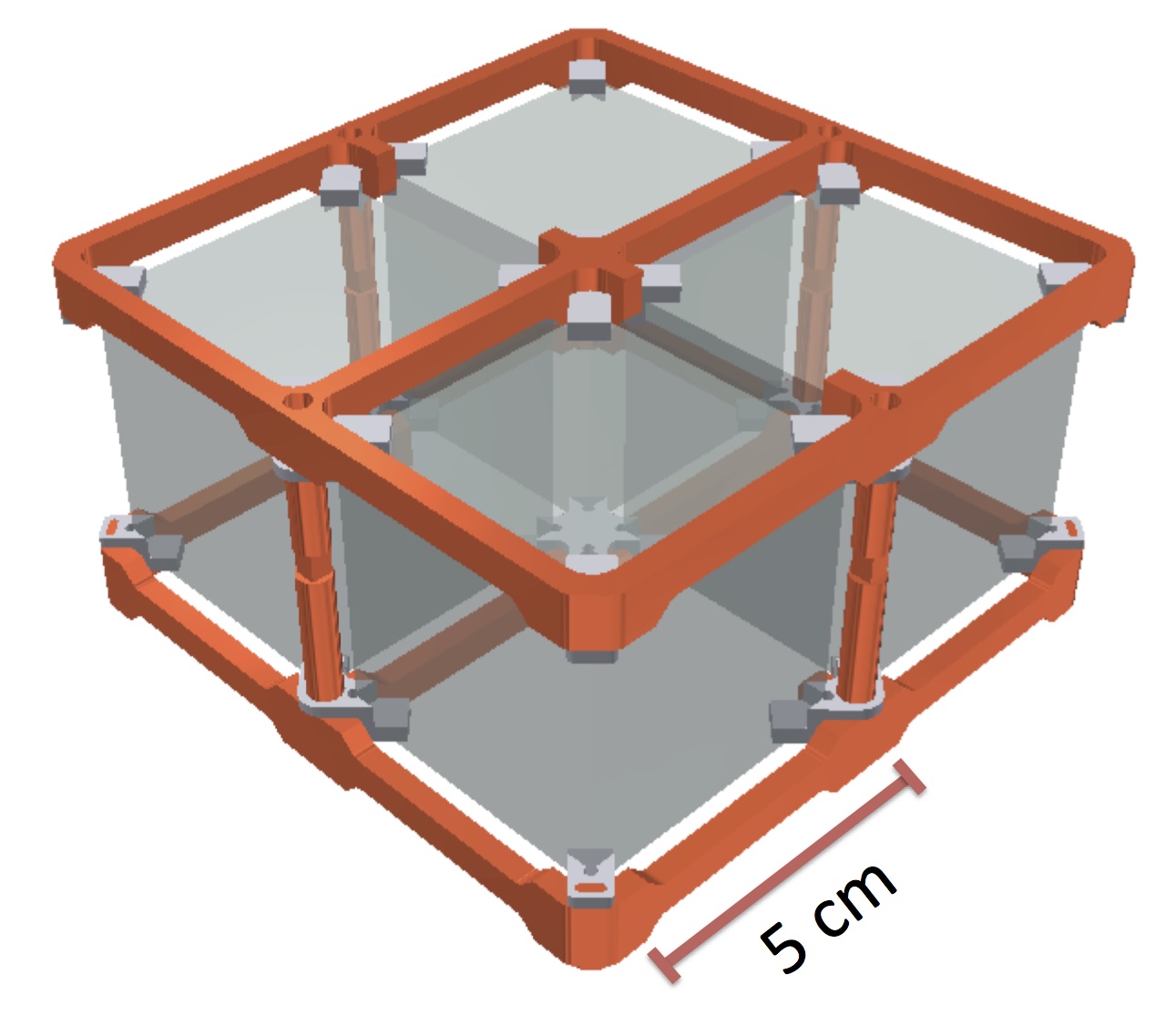}}}\\
 & \multirow{-10}[6]{*}{\subfloat[]{\label{fig:module}\includegraphics[width=0.35\columnwidth]{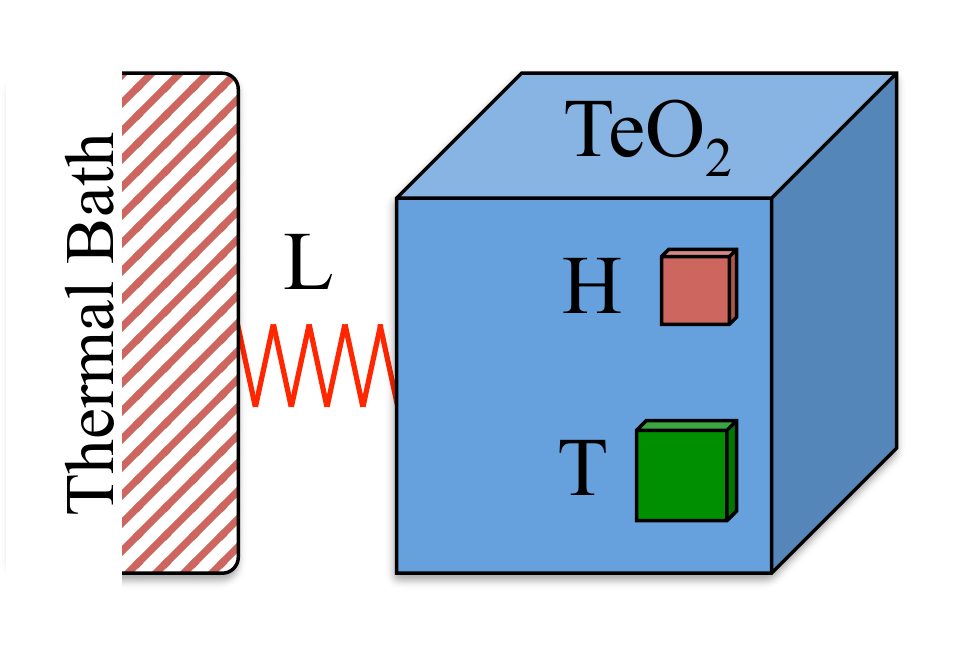}}} \\
 \end{tabular}
\caption{(a)~Illustration of the planned 19-tower CUORE detector array. 
(b)~Closeup of a single tower floor showing four TeO$_2$ crystals held inside their copper frame by PTFE spacers. 
(c)~Schematic diagram of an individual TeO$_2$ crystal bolometer. Each crystal is instrumented with a heater~(H) and a thermistor~(T); the PTFE spacers and sensor readout wires act as weak thermal links~(L) between the crystal and the thermal bath of the copper frame.}
 \end{centering}
\end{figure}

% INTRO, CRYOGENIC BOLOMETERS
The general strategy when searching for $0\nu\beta\beta$~decay is to look for a signature produced by the two final-state electrons, which
would be emitted simultaneously and have a combined energy equal to the decay energy (aka $Q$-value) of the isotope under study.
CUORE will use TeO$_2$ crystals as cryogenic bolometers to search for $0\nu\beta\beta$~decay of $^{130}$Te.
When a TeO$_2$ crystal is cooled to 10~mK, its heat capacity becomes so small that a single particle interaction depositing just a few
keV inside the crystal will produce a measurable rise in its temperature---i.e., the crystal functions as a highly sensitive calorimeter.
The amplitude of the temperature increase is proportional to the energy deposited ($\Delta T/\Delta E\sim 10\1{-}20\1{\mu K/MeV}$), so the basic experimental 
method is to compile an energy spectrum from temperature pulses and look for an excess of events above background at $\sim$~2528~keV, the 
$Q$-value for $\beta\beta$~decay of $^{130}$Te~\cite{Redshaw_Q_2009, Scielzo_Q_2009, Rahaman_Q_2011}.
In this so-called ``source$=$detector'' approach the TeO$_2$ crystal serves a dual role: it contains the decay isotope and also acts as the detector.
This method offers the advantages of high efficiency, scalability, and in our case excellent energy resolution, which is critical to discriminating
any $0\nu\beta\beta$~decay peak in the measured energy spectrum. 
%The primary disadvantage to the approach is that it doesn't permit particle discrimination.

% DETECTOR & CRYSTALS
The CUORE detector will consist of a close-packed array of 988~independent TeO$_2$ crystal bolometers arranged into 19~towers~(Figure~\ref{fig:tower}). 
The basic detector element is a 5$\times$5$\times$5~cm$^{3}$ crystal instrumented with a temperature sensor and a resistive heater. 
Each crystal weighs $750\1{g}$, giving a total detector mass of 741~kg. $^{130}$Te has a natural isotopic abundance of 34.2\%---the highest among 
the $0\nu\beta\beta$~decay candidate isotopes~\cite{Fehr_TeIA_2004}---so the detector will contain 206~kg of source isotope.

The crystals were manufactured from 2009--2013 by the Shanghai Institute of Ceramics, Chinese Academy of Sciences~(SICCAS). 
We worked in close conjunction with SICCAS to develop the rigorous quality and radiopurity controls followed during crystal production~\cite{Arnaboldi:2010fj}. 
To guard against radioactive contamination of the crystal bulk and surfaces we performed high-sensitivity radiopurity checks (e.g., using ICP-MS) 
along the entire production chain, from raw-material synthesis to crystal growth to the final surface treatment. 
Finished crystals were shipped to Italy by sea. Although this took longer than shipping by air, the increased transit time was more than
compensated for by the reduced cosmogenic activation of the crystals from being exposed only to sea-level cosmic-ray flux.

% NTD
Each TeO$_2$ bolometer is instrumented with a neutron-transmutation-doped~(NTD) germanium thermistor that serves as a temperature sensor~\cite{Haller}. %{Itoh_NTD}. 
The device is glued to the crystal surface via a dot matrix of bicomponent epoxy. Previous work by the collaboration has demonstrated that epoxy dots provide a robust thermal 
coupling between the sensor and the crystal while also reducing the mechanical stresses that arise from differences in their thermal contraction rates during cryogenic cooldowns.
(If a continuous veil of epoxy is used instead, the sensor will often detach with a fragment of the crystal during the cooldown step.)
The thermistors were produced by neutron irradiation of pure Ge wafers in a research reactor for precise lengths of time, and then dicing the wafers into 
$\sim3\times3\times1$~mm$^{3}$ chips and sputtering gold contacts on the ends of each chip.
Doped germanium just below the metal-insulator transition region functions as a sensitive, high-resistance thermal sensor with an exponential $R$-$T$ curve, 
$R\approx R_0\exp{(\sqrt{T_0/T})}$~\cite{McCammon:2005yj}, where parameters $R_0$ and $T_0$ depend on the doping density and must be measured experimentally; typical values for the 
NTD thermistors produced for CUORE are $R_0=1\1{\Omega}$ and $T_0=4\1{K}$.
The NTD technique produces uniform doping density over an entire germanium wafer and thus guarantees uniform $R$-$T$ properties and excellent performance 
for all sensors produced from it. 
The CUORE thermistors were characterized in multiple studies utilizing collaboration-managed dilution-refrigerator cryostats, and their bolometric performance was also 
evaluated in test runs (see Sec,~\ref{sec:fwhm}).

% HEATER
The thermal response of a bolometer varies with its temperature, so each crystal is also instrumented with a Joule heater based on 
P-doped Si~\cite{andreotti_production_2012} for the purpose of monitoring the bolometer's performance over the natural temperature
fluctuations~($\Delta T<1$~mK) that occur in the course of normal cryostat operation~\cite{Alessandrello:1998bf}. 
A brief, precise pulse of current is sent through the heater at regular, known intervals, thereby injecting into the crystal a fixed amount of energy 
simulating a $\sim~3\1{MeV}$~event. These reference signals are used later in the offline data analysis to correct for variations in the 
bolometer's thermal gain with time. Each heater consists of a resistive meander obtained by ion implantation on a Si chip.
The device's resistance is $300\pm12\1{k\Omega}$ at 10~mK and is stable to within 0.1\% below 4.2~K.

% FIGURE OF MERIT
A commonly used ``figure of merit'' expression for describing experimental sensitivity to $0\nu\beta\beta$~decay half life is 
\[
T^{0\nu}_{1/2} \propto \eta\cdot a \cdot \sqrt{\frac{M\cdot t}{b\cdot \Delta E}} ~,
\]
where $\eta$ is the physical detector efficiency, $a$ is the isotopic abundance of the $0\nu\beta\beta$~decay candidate, $M$ is the total detector mass, $t$ is its live time, $b$ is the background rate per unit detector mass per energy interval, and $\Delta E$ is the detector's energy resolution~\cite{CUORE_sensitivity_2011}. 
For extremely rare processes like $0\nu\beta\beta$~decay, stringent requirements on energy resolution and background are critical.

%______________________________________________________________
\subsection{Energy resolution of CUORE bolometers} 
\label{sec:fwhm}

\begin{figure}
\begin{center}
\includegraphics[width=0.8\columnwidth]{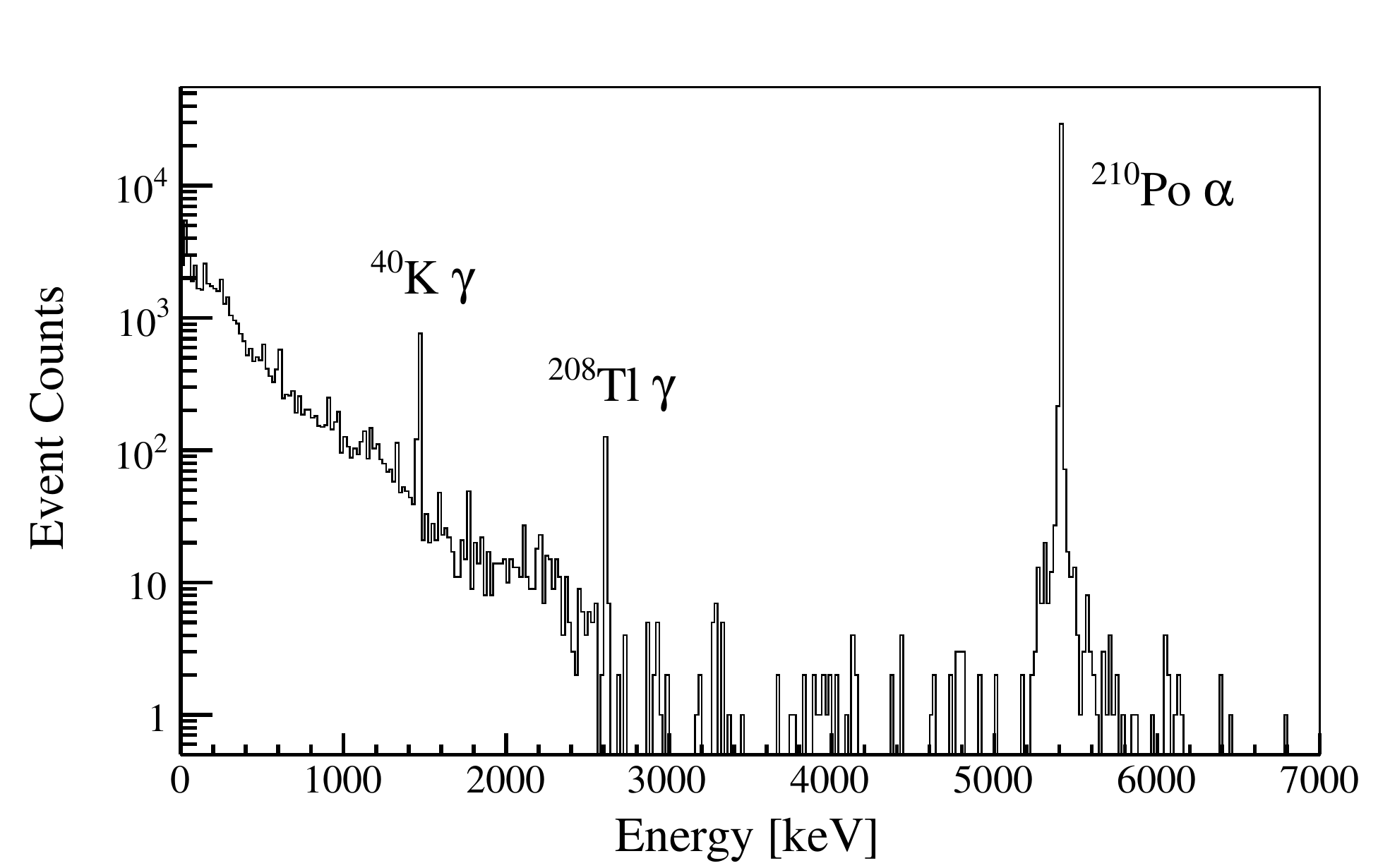}
\caption{Energy spectrum of background events from one CCVR. % CCVR7
The three prominent peaks are the 1461-keV $\gamma$-ray line from $^{40}$K decay, the 2615-keV $\gamma$-ray line from $^{208}$Tl decay, and the 5407-keV $\alpha$ peak at the $Q$-value for $^{210}$Po decay. The $^{210}$Po peak is accompanied by a less prominent peak at 5304~keV which arises because sometimes the decay $\alpha$ is detected but the corresponding nuclear recoil is not---namely, when the decay occurs on a copper surface facing the detectors.
The live time of this CCVR was 21.5~days, corresponding to 0.17~kg$\cdot$y of detector exposure. }
\label{fig:CCVR_E_spectrum}
\end{center}
\end{figure}

Over the course of TeO$_2$ crystal production, the collaboration conducted ten CUORE Crystal Validation Runs~(CCVRs) in which we operated a small fraction of 
the newly manufactured crystals as bolometers in order to check their radiopurity and performance~\cite{CCVR_bkg_2012}. 
These monthlong test runs also served as excellent R\&D platforms for testing post-Cuoricino improvements to the PTFE and copper frame designs, 
for characterizing NTD thermistors, and for studying the energy resolutions of the bolometers at different working temperatures.

For each CCVR we randomly selected four crystals from the latest production batch received at LNGS (typically consisting of $\sim$~100 crystals) and assembled them into a module equivalent to a single floor of a CUORE tower (Fig.~\ref{fig:2by2}). We used CUORE-type copper frames, which have less surface area facing the crystals than those used in Cuoricino in order to reduce $\alpha$-related backgrounds, as well as new PTFE spacers which had been redesigned to reduce vibrational noise. The different thermal expansion coefficients of copper, PTFE, and TeO$_{2}$ have been exploited in such a way that the crystals are held increasingly tighter as the detector temperature decreases. 
Each crystal was instrumented with two NTD Ge thermistors, but the manual work of gluing thermistors to crystals and wiring them to the cryostat was delicate and labor-intensive and as a result we typically lost some channels during detector cooldown.
In some CCVRs each crystal was also instrumented with a heater, but in several runs they were omitted for expediency because the 5407~keV line from $\alpha$ decays of intrinsic, short-lived $^{210}$Po contamination could be used instead to monitor fluctuations in the crystals' thermal gain.

Each CCVR bolometer module was mounted inside a copper canister and cooled down to cryogenic temperatures inside our R\&D cryostat in Hall~C at LNGS.  We reached base working temperatures in the range 12--22~mK, as the dilution refrigerator's performance varied from run to run.
The energy spectrum of background data from a single CCVR is shown in Figure~\ref{fig:CCVR_E_spectrum} as an example. The energy resolution of CCVR bolometers was customarily characterized as the full width at half maximum~(FWHM) of the most populated peak, which is the $\alpha$ line at 5407~keV from $^{210}$Po decay. When the $^{40}$~K $\gamma$~peak was sufficiently populated to permit evaluation we found its energy resolution to be statistically consistent with the $^{210}$Po $\alpha$ peak.

\begin{figure}[tp]
\begin{centering}
 \begin{tabular}{c c}
 \subfloat[]{\label{fig:CCVR_FWHM_Hist}\includegraphics[width=0.45\columnwidth]{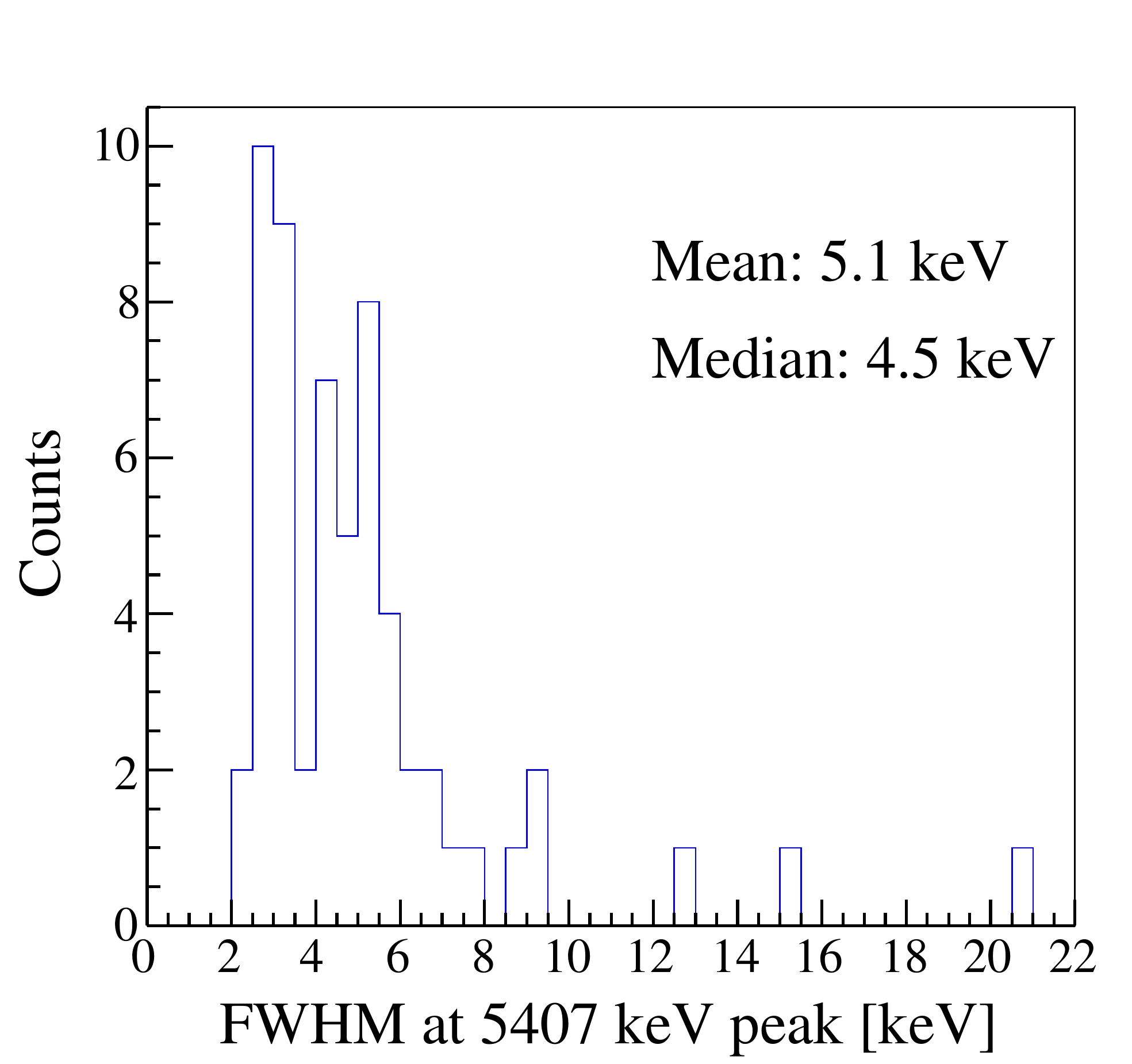}} 
   \hspace{0.05\columnwidth}
\subfloat[]{\label{fig:CCVR_FWHM_Graph}\includegraphics[width=0.45\columnwidth]{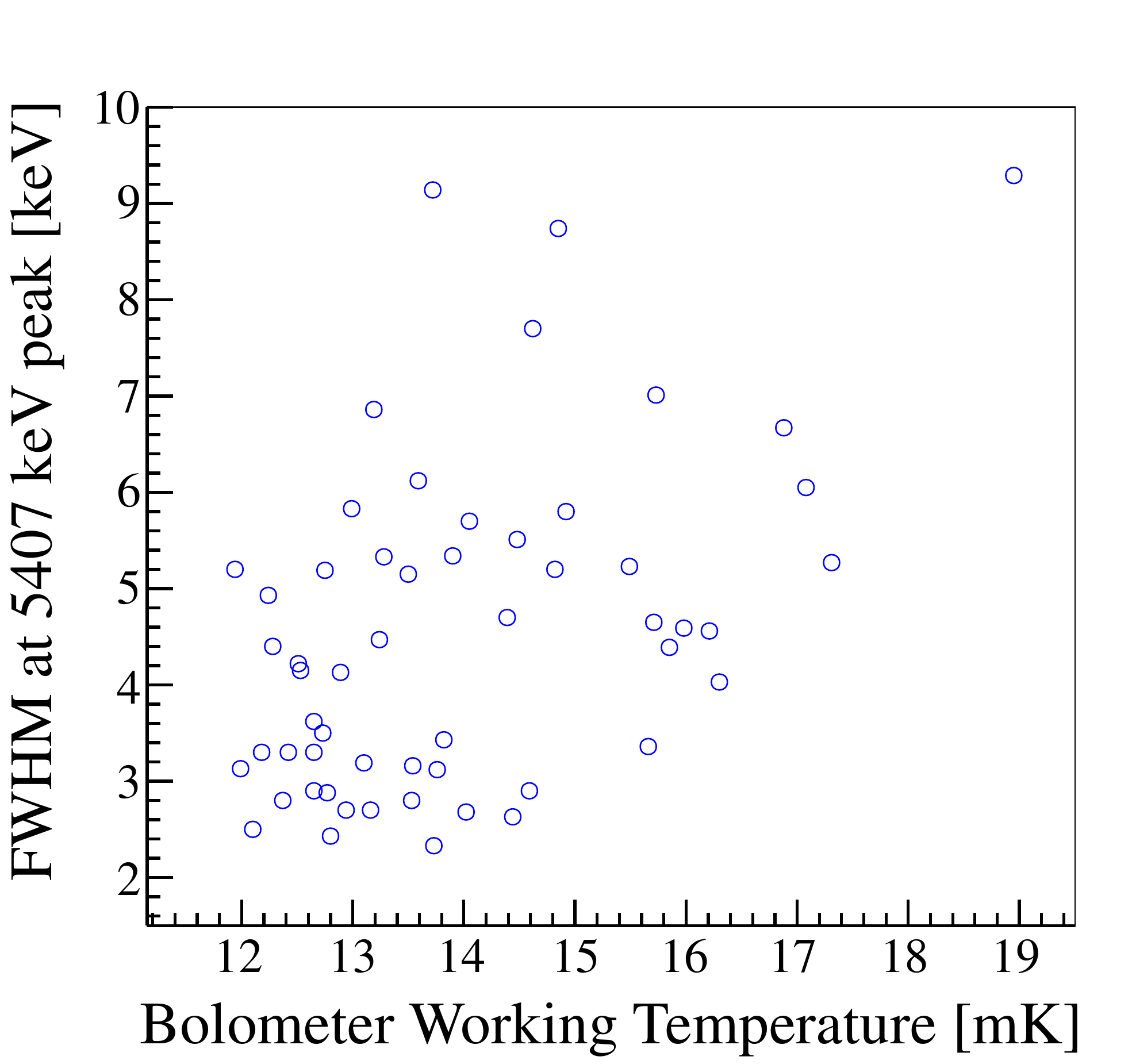}} \\
 \end{tabular}
\caption{(a)~Distribution of energy resolutions at 5407~keV for all CCVR bolometers. (b)~Bolometer energy resolution vs.~detector working temperature for CCVRs 1--9.}
 \end{centering} 
\end{figure}

The distribution of measured energy resolutions at 5407~keV for all CCVR bolometers is shown in Figure~\ref{fig:CCVR_FWHM_Hist}. The average energy resolution of all 69 active channels is 5.1~keV, a result that is noticeably skewed by three noisy detectors with energy resolutions larger than 10~keV FWHM. In Figure~\ref{fig:CCVR_FWHM_Graph} we plot the energy resolution vs.~bolometer working temperature for all except the three worst performing bolometers. 
When the bolometers were operated below 13~mK, close to the CUORE target temperature of 10~mK, we consistently achieved our target energy resolution of 5~keV. It is worth noting that the R\&D cryostat in which the CCVRs were performed is a test facility subjected to frequent modifications that affect its performance, and that its base-temperature rating is $\sim$~2~mK higher than that of the CUORE-0 cryostat and the future CUORE cryostat. We therefore expect bolometer performances in CUORE to be the same or better than what was seen in the CCVRs.

%______________________________________________________________
\subsection{Backgrounds}
\label{sec:bkg}

The paramount concern in $0\nu\beta\beta$~decay searches is suppressing backgrounds that could hide a decay signal. Common sources of background include: cosmic-ray muons and their byproducts, such as cosmogenically activated detector materials; $\gamma$~rays from natural uranium- and thorium-chain radioactivity in the detector, surrounding hardware, and the environment; $\alpha$~particles from surface contamination of materials facing the bolometers; and the irreducible tail from $2\nu\beta\beta$~decay at the end point of the $\beta\beta$ decay energy spectrum. For CUORE, the primary concerns are $\gamma$ rays and $\alpha$ particles from radioactive decays inside the detector and the cryostat. 
The tail from $2\nu\beta\beta$~decay will be negligible, as it will be several orders of magnitude smaller than the other backgrounds due to the excellent energy resolution of the bolometers and the consequently narrow region of interest~(ROI) in the energy spectrum.

% ENVIRONMENTAL/CRYOSTAT
Environmental backgrounds will be strongly suppressed in CUORE through a combination of location and shielding. The underground LNGS host facility is located at an average depth of $\sim 3600$~m water equivalent, which reduces the muon flux from its surface value by roughly six orders of magnitude to $\sim 3\times10^{-8}\1{\mu/cm^2/s}$~\cite{Ambrosio:1995cx, Mei:2005gm, Selvi:2009, Bellini:2012te} and thus dramatically limits background from cosmic muons and muon-induced neutrons and $\gamma$ rays. The primary cosmogenic activation products in TeO$_2$ crystals and copper include $^{60}$Co, $^{110}$Ag, and $^{110m}$Ag, and the estimated background rates in the ROI from those products are at least an order of magnitude smaller than backgrounds coming from surface contamination on the copper facing the crystals~\cite{CUORE_projected_BG}. In order to limit cosmogenic activation of the copper used in the experiment, we store the copper underground and bring it aboveground only when necessary for machining and cleaning~\cite{Alessandria:2012zp}.

The CUORE cryostat will be surrounded by a 73-ton octagonal external shield designed to screen the detector from environmental $\gamma$ rays and neutrons. 
The shield has three layers: an outermost layer consisting of a floor of $\sim$~20-cm-thick 5\%~borated polyethylene~(PE) and sidewalls of a 18-cm-thick pure PE to thermalize and absorb neutrons; a 2-cm-thick side layer of boric-acid powder to absorb neutrons; and an innermost layer of lead bricks of minimum thickness 25~cm to absorb $\gamma$ rays.
% Inside the sidewalls of the lead will be a 0.5-cm copper lining. (Note: Apparently this is an old design and is no longer the case.)

Inside the cryostat two cold lead shields will provide additional protection: a 6-cm-thick layer of ancient Roman lead~\cite{Alessandrello:1998xy}, located between the 4~K and 600~mK copper vessels and thermally anchored to the 4 K vessel, will shield the detectors from radioactivity in the outer vessels and superinsulation, and a 30-cm-thick disc of modern and Roman lead at 0.01~K below the mixing chamber plate will shield the detectors from radioactivity in the overhead cryostat apparatus. The close-packed detector array itself will also provide a measure of passive and active self-shielding, the latter via vetoing of simultaneous events in adjacent crystals. 
%The detector's combined self-shielding capability is responsible for the lower value in the estimated range of upper limits on the surface-related background.

From Monte Carlo simulations we find the expected environmental muon, neutron, and $\gamma$ background rates in the ROI to be $(1.04\pm 0.22)\times10^{-4}$, $(8.56\pm 6.06)\times10^{-6}$, and $<3.9\times10^{-4}$~counts/keV/kg/y (90\%~C.L. upper limit), respectively~\cite{Andreotti:2009dk, CUOREExternal}. These values are orders of magnitude smaller than the $\alpha$ and $\gamma$ backgrounds expected to come from the experimental apparatus itself.

% DETECTOR BULK CONTAMINATION
The 2528-keV $Q$-value for $0\nu\beta\beta$~decay of $^{130}$Te lies above most naturally occurring $\gamma$ backgrounds except the 2615-keV line from $^{208}$Tl. Consequently, the $\gamma$ background in the CUORE ROI should come almost entirely from $^{208}$Tl present in the experimental setup. The upper limits on parent $^{238}$U and $^{232}$Th bulk contaminations in the TeO$_2$ crystals,
as determined from CCVR measurements, are $6.7\times10^{-7}$ and $8.4\times10^{-7}$~Bq/kg, respectively, which translate to $<10^{-4}$~$\gamma$~counts/keV/kg/y in the ROI~\cite{CCVR_bkg_2012}.
The $\gamma$ background due to bulk contamination in the experimental setup should be less than $6\times10^{-3}$~counts/keV/kg/y~(90\% C.L.)~\cite{CUORE_projected_BG}.

% DETECTOR SURFACE CONTAMINATION
The largest background in the ROI is expected to come from $\alpha$ particles emitted by contaminants on the surface of the copper in the detector towers and the innermost thermal shield. We tested four techniques for minimizing effects from copper surface contamination~\cite{Alessandria:2012zp, Parylene}. Two of the techniques involved complex procedures for removing contamination by etching a thin layer off the copper surface, while the other two techniques involved covering the copper with a material (either polyethylene wrapping or a parylene conformal coating) to absorb emitted $\alpha$ particles and thereby prevent them from impinging on the crystal detectors. Based on the test results and practical concerns, we elected to subject all copper components facing the bolometers---i.e., both the parts in the towers and the innermost thermal shield---to a surface-cleaning process consisting of abrasive tumbling, electropolishing, chemical etching, and magnetron plasma etching~\cite{Alessandria:2012zp}. Recent results from the currently running CUORE-0 detector (see Sec.~\ref{sec:q0}) indicate the copper surface cleaning has been effective. The expected upper limit on surface-related backgrounds in the ROI in CUORE, extrapolated from test results and the recent performance of CUORE-0, 
is 1--2$\times10^{-2}$~counts/keV/kg/y~\cite{CUORE_projected_BG}.

% !TEX root = CUORE_status_review_2014.tex
\section{CUORE prototype: CUORE-0}
\label{sec:q0}

\begin{figure}[tp!]
\begin{centering}
 \begin{tabular}{ccc}
 \subfloat[\Qz{} tower]{\label{fig:q0tower}\includegraphics[height=0.7\columnwidth]{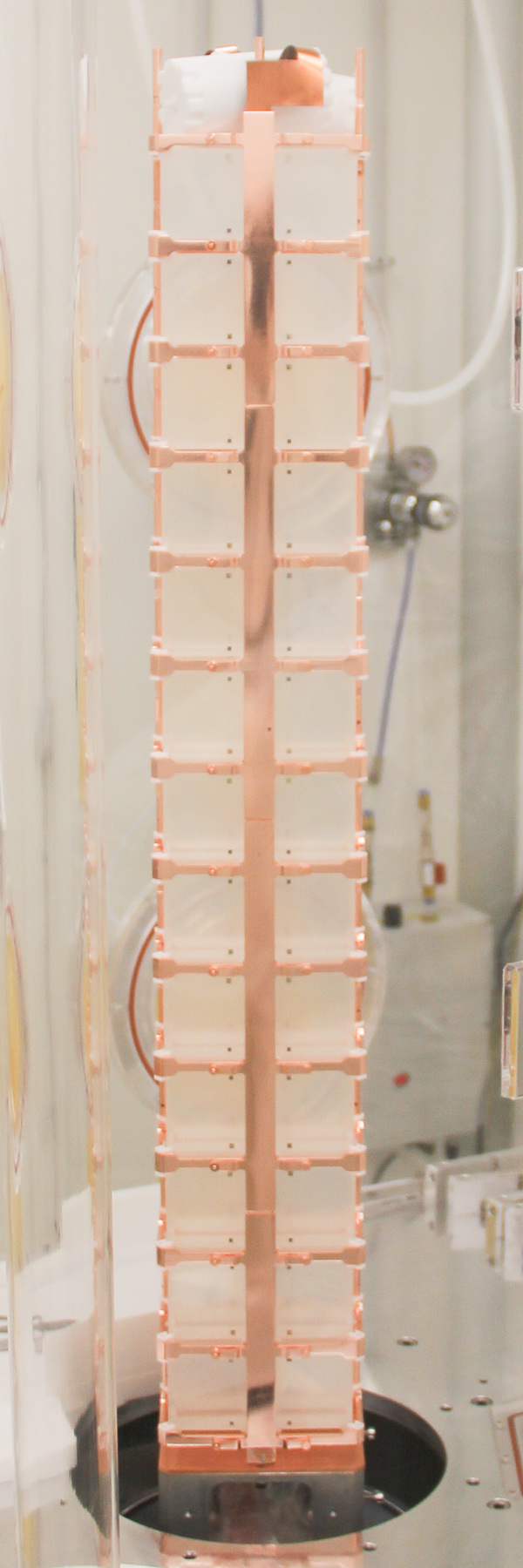}} 
  \hspace{0.1\columnwidth}
\subfloat[\Qz{} temperature]{\label{fig:q0T}\includegraphics[height=0.7\columnwidth]{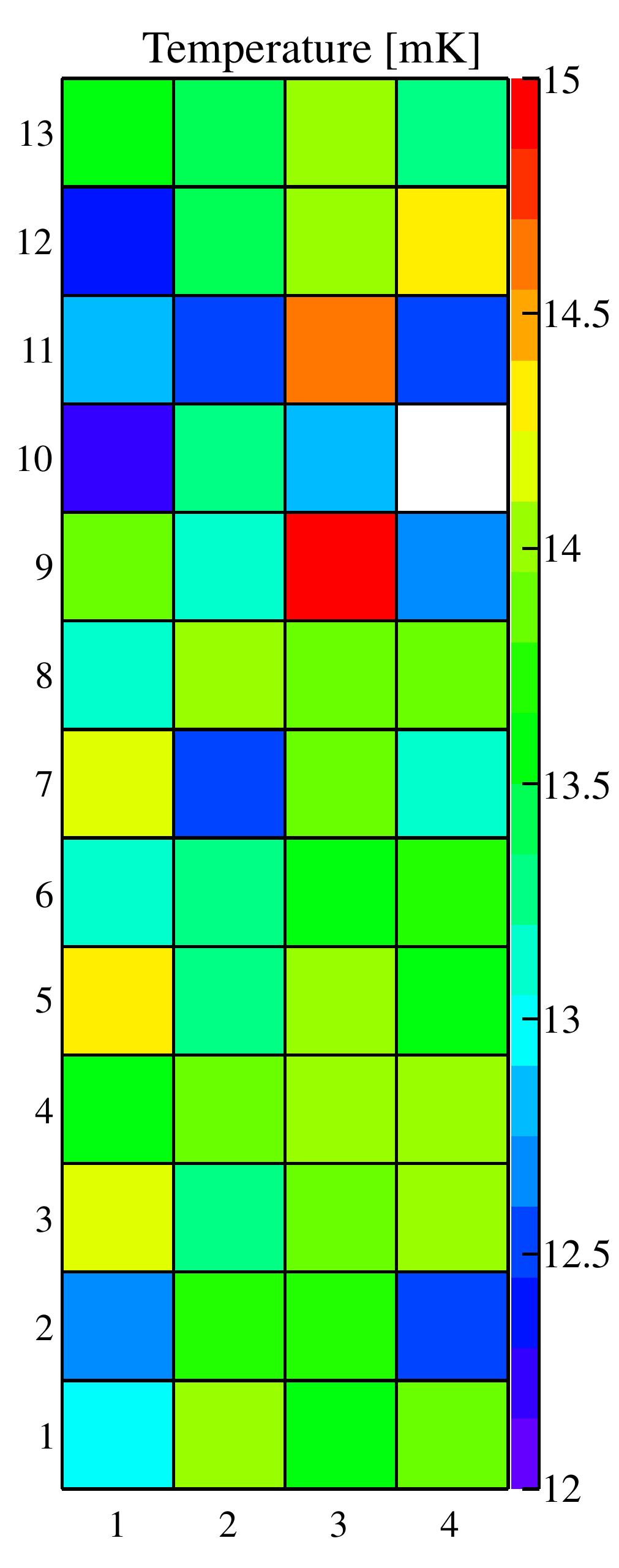}} 
 \hspace{0.1\columnwidth}
\subfloat[\Qz{} $E$ resolution]{\label{fig:q0FWHM}\includegraphics[height=0.7\columnwidth]{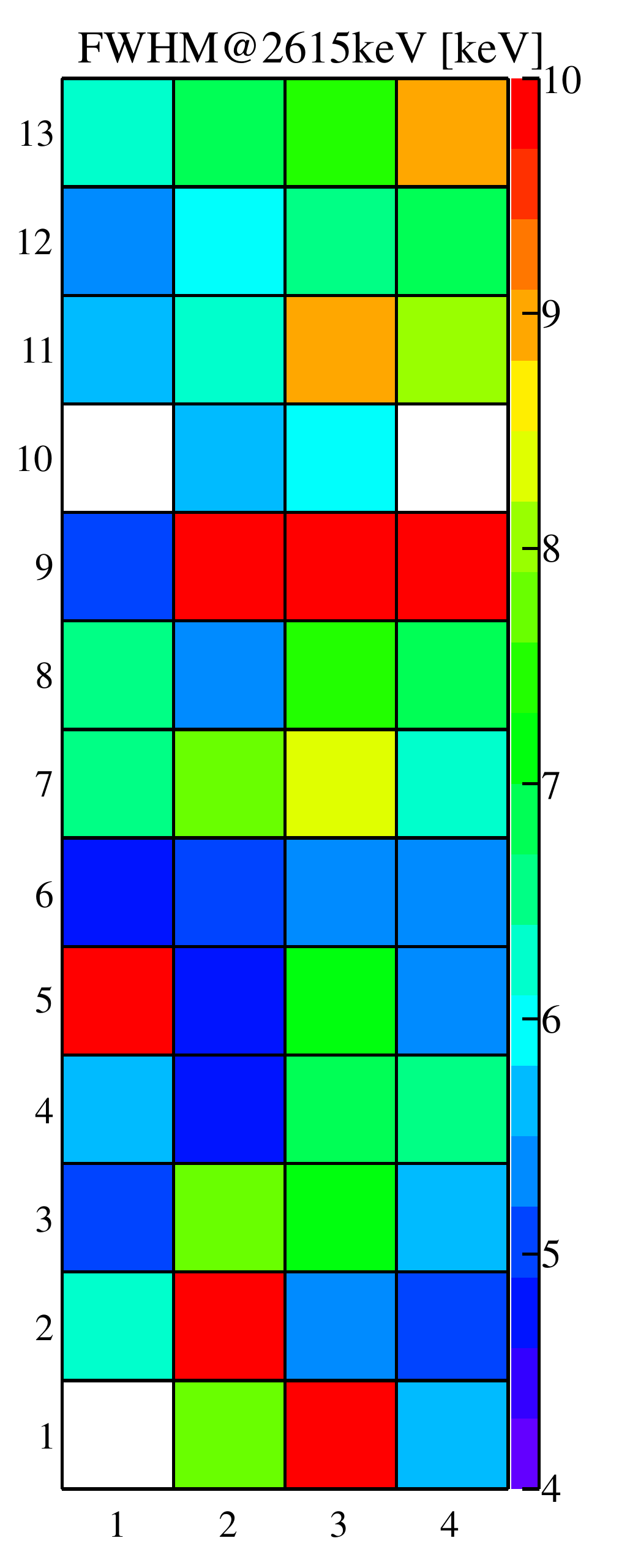}} \\
 \end{tabular}
\caption{(a)~The \Qz{} tower inside a glove box during its assembly.
(b)~Base temperatures of the functioning \Qz{} crystals, as calculated from the measured resistances of their thermistors in conjunction with 
$R_0$ and $T_0$ values previously determined in thermistor characterization studies (see Sec.~\ref{sec:bolometers}).
It should be mentioned there is good evidence that a thermistor's $R$-$T$ behavior can be modified by mechanical stresses caused by its 
glue-spot connections to the crystal, and this effect could be responsible in part for the seemingly erratic distribution of detector temperatures.
(c)~Energy resolution~(FWHM) of each detector channel, determined from calibrations performed regularly during March--September 2013.
The mean energy resolution in the calibration data is 6.8~keV, and the median is 6.0~keV.}
\end{centering}
\end{figure}

In order to commission the CUORE detector assembly line, confirm the effectiveness of our copper-surface-cleaning technique, and validate post-Cuoricino
improvements to the tower design, we built \Qz{}, a single CUORE-type tower containing 52~TeO$_2$ bolometer modules~(Figure~\ref{fig:q0tower}), and have been operating it since March 2013~\cite{CUORE0Paper}. \Qz{} is the first tower produced using CUORE assembly techniques and materials, including surface-cleaned copper. The detector's total mass is 39~kg, with 11~kg of $^{130}$Te isotope. In this section we review the construction and commissioning of CUORE-0, its performance and background measurements, and its potential physics reach.

%______________________________________________________________
\subsection{Construction and commissioning}

The \Qz{} tower was constructed according to standard CUORE detector assembly procedures, described in Section~\ref{sec:assembly}. After being built, the tower was enclosed in a copper thermal shield and installed in the former Cuoricino cryostat. 
\Qz{} therefore shares much of the same infrastructure used in Cuoricino, such as the external shielding, the Faraday cage, and the data acquisition~(DAQ) 
hardware~\cite{Arnaboldi:2008ds}.
A Plexiglas shield surrounding the cryostat is continuously flushed with nitrogen gas to prevent ingress of radon.  
% The tower was kept under nitrogen atmosphere during all of these operations.
We operate \Qz{} at $\sim13\1{mK}$ due to the limitations of the aged cryostat.

The base temperature of each bolometer can be calculated from the measured resistance of its NTD thermistor (Figure~\ref{fig:q0T}). Of the 52 bolometer channels, one is not functional due to a failed wire bonding to its thermistor during assembly. In addition, one heater could not be bonded during assembly, and the connection to another heater was lost during the initial detector cooldown. We considered this situation acceptable and proceeded with data taking.

The DAQ hardware includes front-end preamplifiers, six-pole low-pass Bessel filters, and high-precision 18-bit National Instruments digitizers operating at 125~S/s~\cite{Arnaboldi:2004jj, Arnaboldi:2010zz}. On the software side we use the \textsc{Apollo} suite developed for CUORE. 
We record both the continuous data stream and software-triggered data samples; each bolometer module is triggered independently with a threshold in the range 50--100~keV. The detectors are calibrated using gamma lines from two thoriated tungsten strings which are lowered into guide tubes between the cryostat and external lead shield once per month. 
To calibrate the bolometers' response across the measured energy spectrum, we use a third-order polynomial to fit the locations of the source-generated gamma peaks in the range 511--2615~keV.

%______________________________________________________________
\subsection{Detector performance and background measurement}
\label{ssec:q0-bg}

\begin{figure}[t]
\begin{center}
\includegraphics[width=0.8\columnwidth]{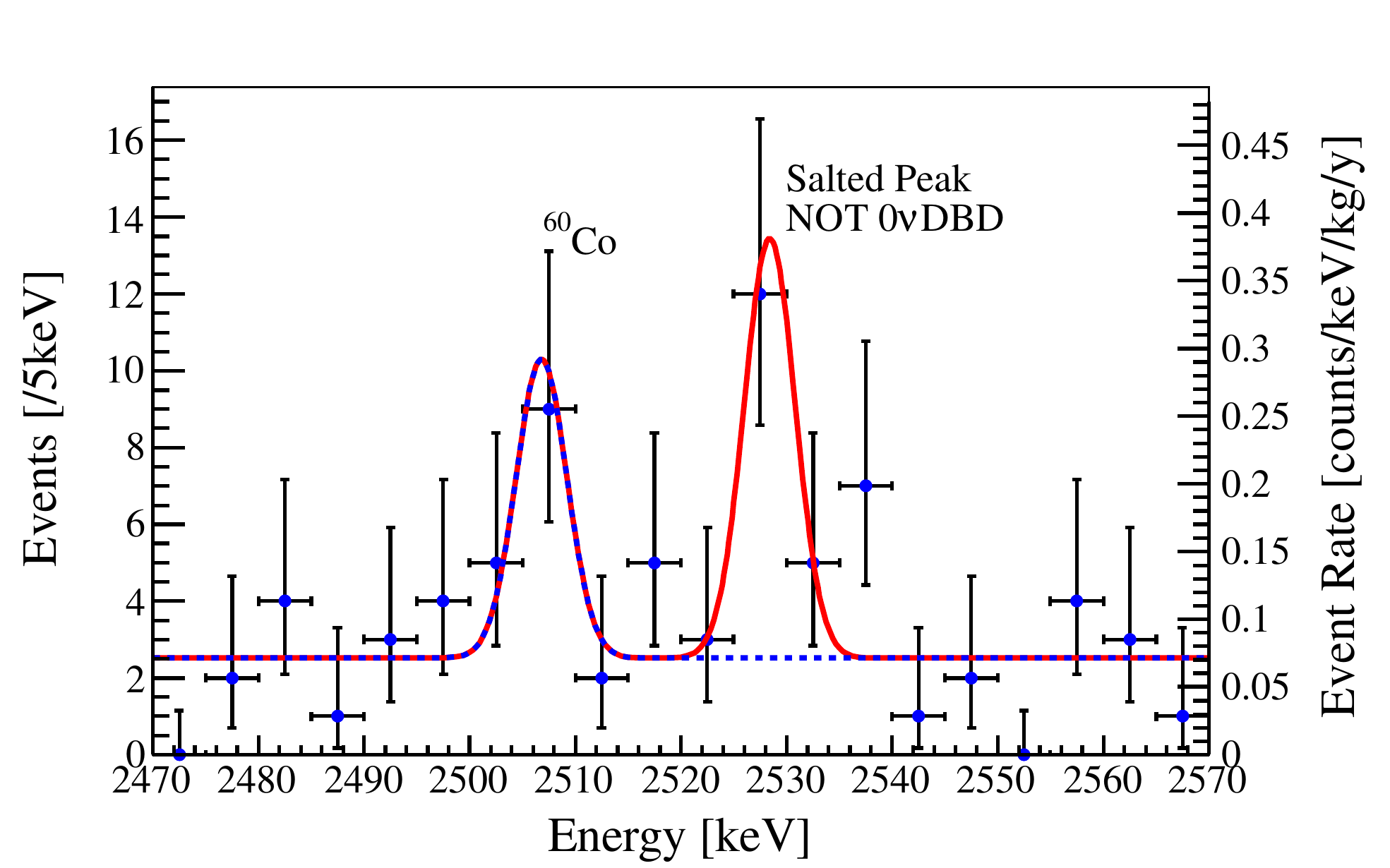}
\caption{CUORE-0 background spectrum in the ROI, with a salted peak at the $Q$-value for $0\nu\beta\beta$~decay of $^{130}$Te. 
The solid red line shows the result of an unbinned maximum-likelihood fit to the full spectrum, while the dotted blue line indicates 
the fit's background component, consisting of the ${}^{60}\n{Co}$ peak and a uniform continuum. 
The fit value for the continuum background is ${0.071 \pm 0.011\1{counts/keV/kg/y}}$. (From~\cite{CUORE0Paper}).}
\label{fig:q0ROI}
\end{center}
\end{figure}

\begin{figure}[tp]
\begin{center}
\includegraphics[width=0.8\columnwidth]{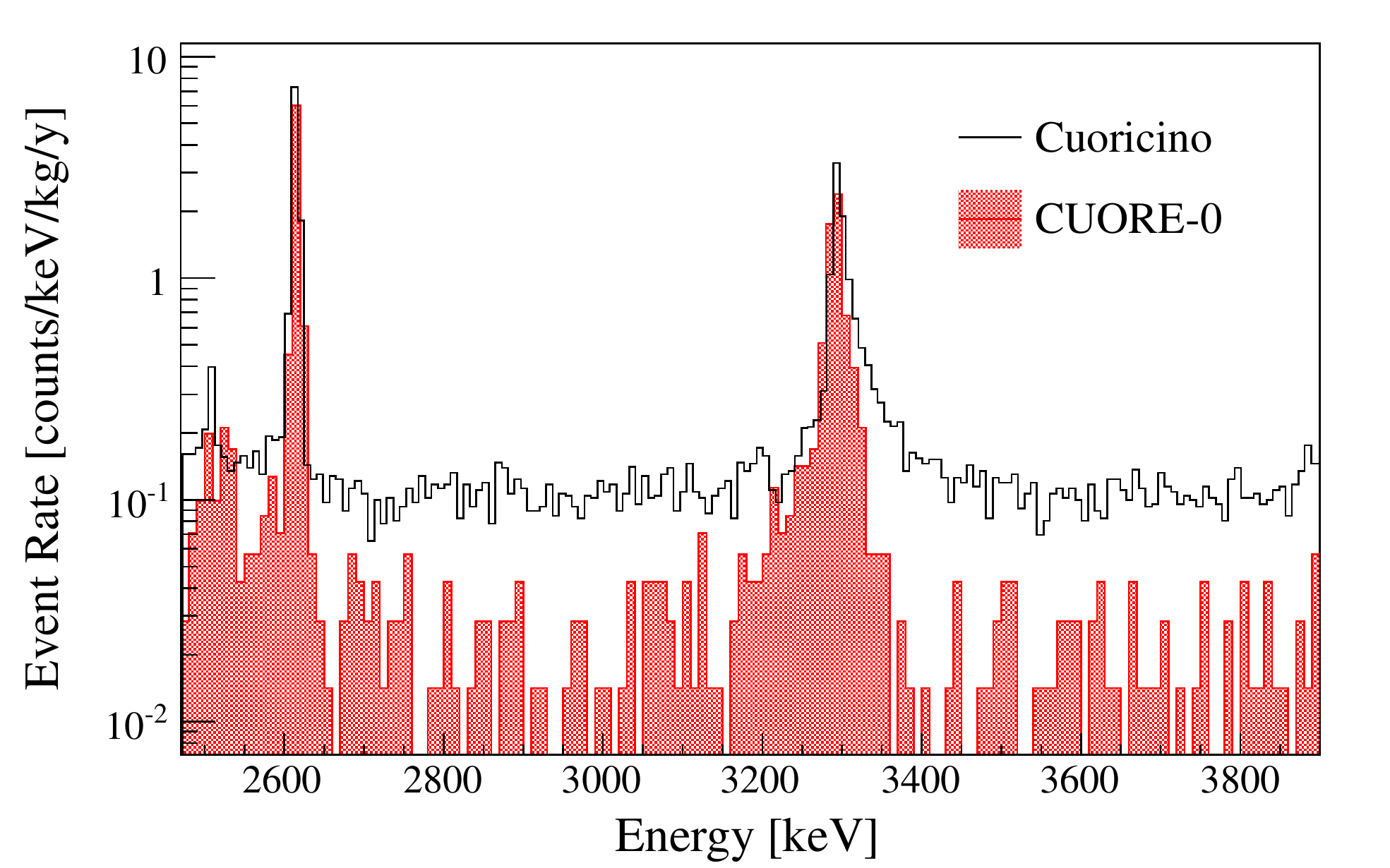}
\caption{Background spectra from CUORE-0~(shaded, red) and Cuoricino~(black) in the energy region dominated by degraded $\alpha$ particles~\cite{CUORE0Paper}.
The sixfold reduction in background achieved by CUORE-0 in the energy ranges 2.7--3.1~MeV and 3.4--3.9~MeV is clearly evident.
The prominent peak at 3.3~MeV is from decay of $^{190}$Pt, which is present in the TeO$_2$ crystals from their growth. 
The non-Gaussian shape of the peak is due to the fact these spectra are made by summing the contributions from many detector channels
having different (Gaussian) energy resolutions and positions, the variations in the latter arising because of calibration uncertainties.}
\label{fig:q0alpha}
\end{center}
\end{figure}

The offline analysis of \Qz{} data follows a standard procedure~\cite{CUORE0Paper} originally developed for the Cuoricino experiment~\cite{Arnaboldi:2008ds, Cuoricino_NDBD_2011}. 
The heater pulses which periodically inject fixed amounts of energy into the bolometers are used to correct for small shifts in thermal gain due to temperature fluctuations.
The amplitudes of the bolometer temperature pulses are extracted via optimum filtering~\cite{gatti_OF_1986} and then converted into energies using calibration data.
Events occurring within $\pm100$~ms of each other in multiple crystals are rejected to reduce background.
Figure~\ref{fig:q0FWHM} shows the distribution of the full width half maximum~(FWHM) values of the 2615~keV $^{208}$Tl decay peak in calibration data
for all 49~active channels with functional heaters used in the data analysis.
At the conclusion of the first phase of data taking the summed exposure of the individual detectors was 7.1~kg$\cdot$y. 
We evaluate the overall detector energy resolution in the non-calibration data to be 5.7~keV, based on the FWHM of the 2615~keV peak in the energy
spectrum created by summing the data for all active channels.

At present the CUORE-0 data in the ROI for $0\nu\beta\beta$~decay of $^{130}$Te is blinded while we accumulate more statistics and work on optimizing event selection.  
To perform the blinding we exchange a random fraction of events within $\pm10$~keV of the decay's $Q$-value with events within $\pm10$~keV of the 2615~keV $\gamma$ peak; the number and identity of the exchanged events are kept secret from the analyzers.
Since the number of events in the $\gamma$ peak is significantly larger than that in any possible $0\nu\beta\beta$ peak, this blinding procedure generates an artificial peak centered at the $Q$-value which hides any $0\nu\beta\beta$~decay signal. 
The $\pm10$~keV exchange width was chosen because it is approximately twice the FWHM energy resolution of the detectors.
Figure~\ref{fig:q0ROI} shows the so-called ``salted peak'' and the nearby $^{60}$Co $\gamma$ peak.

To find the average background rate in the ROI, we use an unbinned maximum likelihood fit in which the likelihood function includes two Gaussians (for the $^{60}$Co and $0\nu\beta\beta$~decay peaks) and a constant continuum which incorporates the $\alpha$ and $\gamma$ backgrounds. The fitted background rate is $0.071\pm0.011$~counts/keV/kg/y.
The main background contributions in the ROI are $\gamma$~rays from decay of $^{208}$Tl coming from $^{232}$Th in the cryostat,
and $\alpha$ particles from radioactive decays on the surface of the detector materials. The former is expected to be similar to the $\gamma$ background measured in Cuoricino at 0.05--0.06~{counts/keV/kg/y}, while the latter can be extrapolated from the measured background rate at higher energies in the range 2.7--3.9~MeV. 
Any deviation from a constant (i.e., flat) continuum background is contained in the systematic error of the fitted background rate.

The $\alpha$ continuum from 2.7--3.9~MeV (excluding the peak in the range 3.1--3.4~MeV from decay of $^{190}$Pt in the crystals) is above all naturally occurring $\gamma$ lines, so the background in that region mainly comes from $\alpha$ particles whose energy has been degraded. In Figure~\ref{fig:q0alpha} we compare \Qz{} with Cuoricino in the $\alpha$ continuum region and find the \Qz{} background rate is $0.019\pm0.002\1{counts/keV/kg/y}$, a factor of six less than in Cuoricino, $0.110\pm0.001\1{counts/keV/kg/y}$.

%______________________________________________________________
\subsection{Projected sensitivity}

\Qz{} data taking is ongoing and expected to continue until CUORE comes online in early 2015. With its improved background compared to the previous generation of bolometer experiments, \Qz{} has the potential to make a significant improvement on the limit for $0\nu\beta\beta$ of $^{130}$Te. With roughly one year of live time, \Qz{} should surpass the half-life limit on $0\nu\beta\beta$~decay of $^{130}$Te established by Cuoricino at $2.8\times10^{24}\1{y}$ (90\% C.L.)~\cite{CUORE0Paper}.

% !TEX root = CUORE_status_review_2014.tex

\section{CUORE status}
\label{sec:status}

In parallel with CUORE-0 data taking, we have also been busy building the CUORE detectors and experimental setup in Hall~A at the LNGS underground facility. This work is scheduled to continue until the end of 2014, with the goal of turning on the experiment in early 2015.

\subsection{Detector assembly}
\label{sec:assembly}

% ASSEMBLY OVERVIEW
Construction of the 19~detector towers is a lengthy, delicate activity that demands a sizable share of the collaboration's attention and resources.
The assembly process is divided into four stages: gluing of thermistors and heaters to crystals, physical assembly of instrumented crystals into a tower, attachment of readout cables to the tower, and wire bonding of the crystals' chips to the readout cables. To minimize oxidization and contamination (especially by radon~\cite{Clemenza:2011zz}), all of these operations are carried out using clean tools inside nitrogen-flushed glove boxes in a dedicated clean room on the first floor of the CUORE hut. Completed towers are stored in nitrogen-flushed canisters to await future installation in the cryostat all at once.

% CRYSTAL GLUING
The gluing of semiconductor chips to crystals is performed inside a dedicated glove box by a semi-automated robotic system to achieve precise and uniform results. First, a cartesian robot equipped with a pneumatic gun dispenses matrices of uniformly sized dots of Araldite Rapid bicomponent epoxy on an upturned thermistor and heater placed atop a precision positioning device. Before the epoxy dots begin to cure, a robotic arm fetches a crystal and places it on a cradle above the chips; an actuator then immediately lowers the crystal to a position where it is separated from the chips by $50\1{\mu m}$. The crystal is left to cure undisturbed for a minimum of 50~minutes before being removed from the positioning device with its newly attached chips. For quality control purposes we take pictures of the epoxy dots before and after the chips are attached to the crystal. Crystal gluing is a near-continuous activity and typical system throughput during normal operation is 6~crystals/day, or roughly one tower's worth of crystals every two weeks. Finished crystals are kept in vacuum-sealed containers inside nitrogen-flushed storage cabinets to await assembly into towers.

% MECH ASSEMBLY
All subsequent tower-assembly operations are performed at a workstation containing a nitrogen-flushed storage garage and a work surface that can host a series of task-specific glove boxes and tools.
In order to increase operational efficiency we generally try to assemble towers in batches of 3--4 at a time.

The first task is to physically assemble chip-equipped crystals, specially treated ultraclean copper pieces~\cite{Alessandria:2012zp}, and PTFE spacers---almost 500 separate parts in all---into a tower. The tower is built one floor at a time, descending into the storage garage as it grows in size.

% STRIPS
Once a tower is built, the next step is to install two sets of flexible printed circuit board~(PCB) cables on opposite sides of it to provide the electrical connections to the cryostat wiring. The cables, which consist of wire traces etched from copper sheet on polyethylene naphthalate (PEN) substrate, are 2.4~m in length to run from the bottom floor of the tower up to the cryostat's mixing-chamber plate. The readout traces terminate in bonding pads located on horizontal arms extending from either side of the readout cables at each tower floor. We first glue the flexible PCB cables to a rigid copper backing using Araldite Standard bicomponent epoxy, and after curing overnight the cable assembly is affixed to the tower frame.

% BONDING
The last step is to connect the crystals' semiconductor chips to the PCB cable traces with 25-$\1{\mu m}$ gold wires. This is accomplished using a modified Westbond 7700E manual wire bonder which has been oriented vertically and mounted on motor-driven rails to enable precise horizontal motion. The 8-mm difference in depth between the chip pads and the copper-trace pads is beyond the wire bonder's reach, so the horizontal rails extend the access depth and make bonding possible. Each gold wire is first ball bonded to a chip pad and then wedge bonded to a copper pad, and then the wedge bond is reinforced with a security ball bond. Two wires are bonded for each electrical connection to provide redundancy. After bonding work on a tower is complete, protective copper covers are installed over the PCB cables and the finished tower is placed inside a nitrogen-flushed storage canister.

CUORE tower assembly began in January 2013, and the full complement of 19 towers was completed in Summer 2014.

\subsection{Cryogenics and calibration systems}

\begin{figure}[t!]
 \begin{tabular}{lr}
 \subfloat[]{\label{fig:cryostat}\includegraphics[height=0.63\columnwidth]{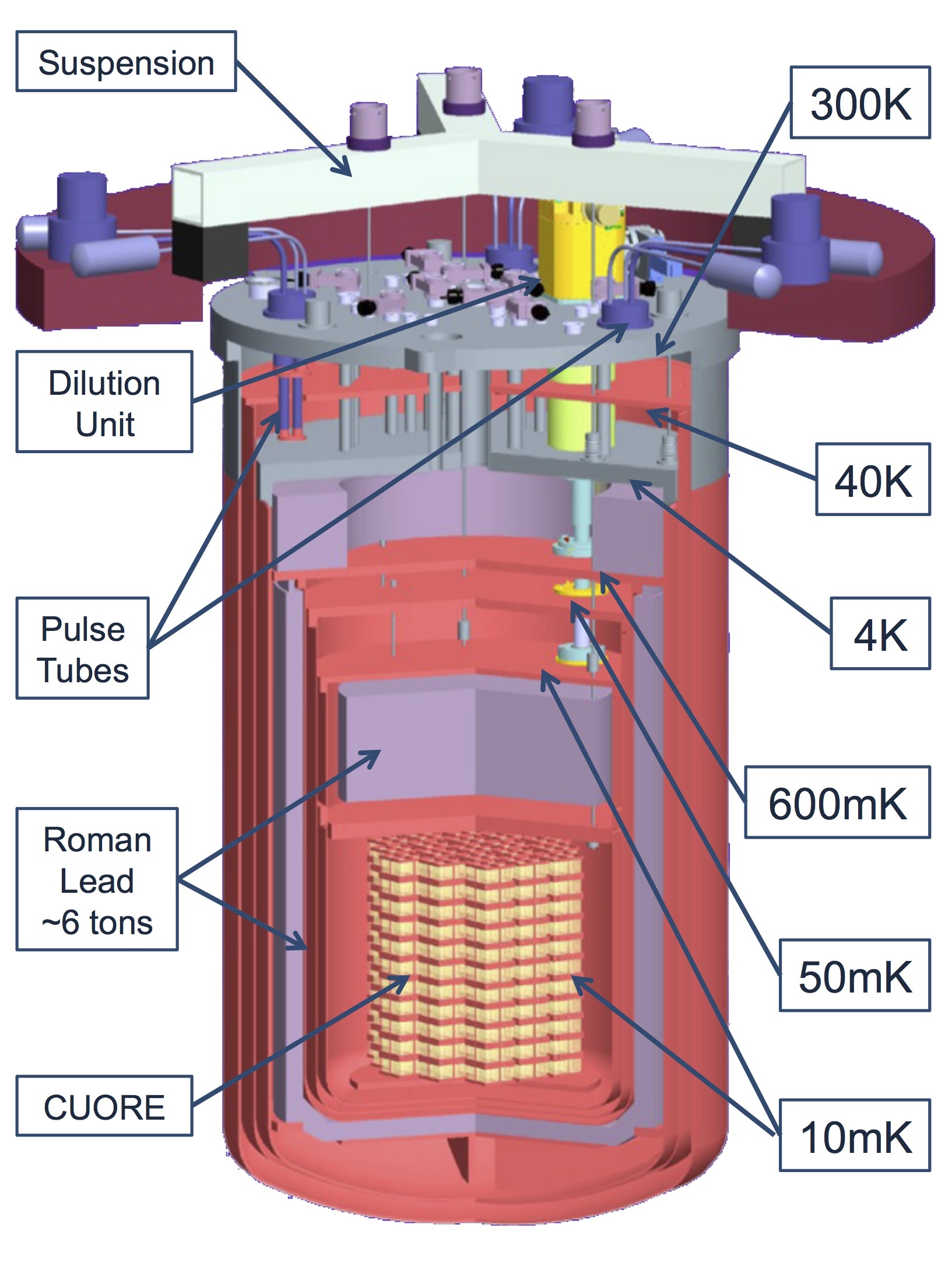}} 
 \hspace{0.1\columnwidth}
 \subfloat[]{\label{fig:dcs}\raisebox{0.1in}{\includegraphics[height=0.6\columnwidth]{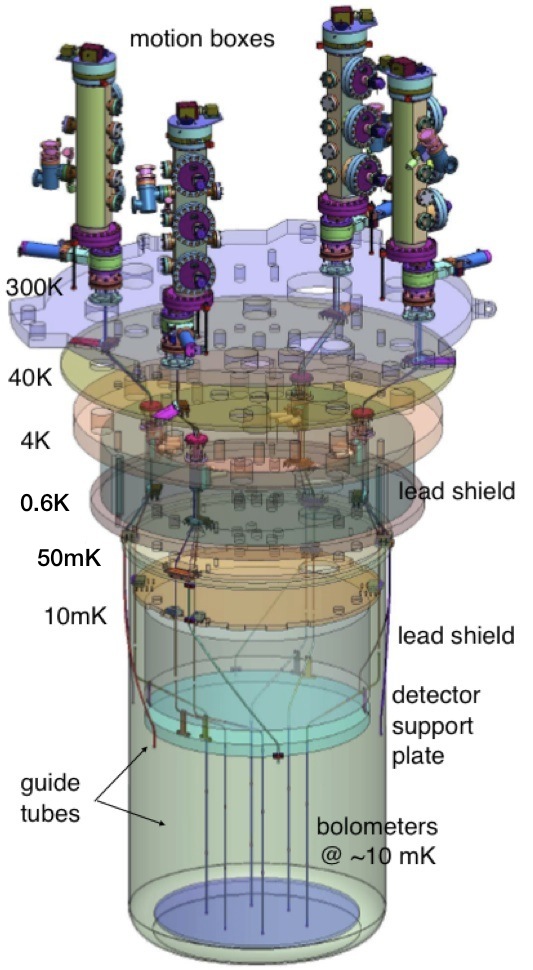}}}\\
 \end{tabular}
\caption{(a)~CUORE cryostat, with major components highlighted. (b)~Detector calibration system~(DCS).}
\label{fig:apparatus}
\end{figure}

The other major challenges in building CUORE are constructing the cryogenics and calibration systems, which will comprise large, complex, interconnected parts packed close together and operating under extremely cold conditions. Given the relatively long time ($\sim$~1~month) needed to close the cryostat and cool the detectors to base temperature, as well as the anticipated five-year running time for the experiment, it is essential that all cryostat systems be carefully designed for robust performance.

The cryogenics system (Fig.~\ref{fig:apparatus}a) encompasses the cryostat and a cryogen-free cooling system, the latter comprising five pulse-tube coolers and a dilution-refrigerator unit~\cite{Cryostat_overview_2012}. 
The cryostat will consist of six nested copper vessels at 300~K (Outer Vacuum Chamber), 40~K, 4~K (Inner Vacuum Chamber), 0.6~K (Still), 0.05~K (Heat Exchanger), and 0.01~K (Mixing Chamber). Construction of all six vessels is complete. We are taking a phased approach to commissioning the cryostat, having started with the outer three vessels instrumented with three pulse tubes~\cite{Cryostat_4K_2014}. We cooled down this partial system twice, successfully reaching 3.5~K on the 4~K plate on the second attempt. The three inner vessels have been cleaned and delivered to LNGS and are presently being installed. To reduce vibrational noise, the detector will be suspended from a Y-beam whose supports are decoupled from the surrounding building structure and the cryostat.

In parallel with cryostat commissioning, the dilution unit~(DU)---a custom-built, closed-cycle, high-power $^{3}$He/$^{4}$He dilution refrigerator---was 
characterized in its own custom test cryostat. The DU was delivered to LNGS in Summer 2012 after passing in-house benchmarking at Leiden Cryogenics. 
It reached 5~mK base temperature at LNGS with a cooling power of $5\1{\mu W}$ at 12~mK. During stable cryostat operation the DU and a subset of the 
five pulse-tube coolers will provide enough cooling power to maintain the detector at base temperature. However, those devices do not have 
sufficient power to cool down the multi-ton apparatus from room temperature in a reasonable time. For this reason the DU and pulse tubes will be 
supplemented during cooldowns with a fast-cooling system---namely, a forced He gas circulation system designed to improve the thermal exchange 
inside the cryostat and thereby reduce the cooling time to $\sim$~one month.

The detector calibration system~(DCS) will be used to lower 12~radioactive source strings under their own weight through a set of guide tubes from the 300~K flange into the 10~mK detector region for the purpose of monthly energy calibrations. The DCS consists of a computer-controlled vacuum motion system above the 300~K flange, a thermalization mechanism at the 4~K flange, and guide tubes which snake through the cryostat's interior and run down between the detectors (Fig.~\ref{fig:apparatus}b). The sources are copper-covered capsules of thoriated tungsten crimped at intervals along a Kevlar string and coated with PTFE to minimize friction. A key challenge is developing a robust system that will not exceed the stringent heat-load constraints of the successive temperature stages inside the cryostat during the insertion and retrieval of source strings. During a test cooldown of the outer cryostat to 4~K we successfully tested a complete calibration source deployment unit operating two strings.

All of the copper components that will be cooled to base temperature---the 10~mK plate and vessel, the tower frames, the tower-suspension
plate, and the DCS~tubes---are made of radiopure electrolytic tough pitch copper alloy~\cite{Alessandria:2012zp,Cryostat_overview_2012}. 
All other copper components in the cryostat are made of oxygen-free electrolytic~(OFE) copper alloy~\cite{Cryostat_overview_2012,Cryostat_4K_2014}.

Electrical signals from the detectors will be carried up to the coldest stage of the cryostat by PCB wires, and then to the outside world via 
NbTi wires running between the Mixing Chamber plate~(0.01~K) and the top of the cryostat~(300~K).
The wires will be arranged in six bundles inserted through six 40-mm inner bore access ports placed between the 300~K and 4~K flanges. 
The wires will be cooled only by radiation inside the bore holes; below 4~K the wires will be cooled by conduction through thermalization
clamps connected to each cold stage of the refrigerator.

% !TEX root = CUORE_status_review_2014.tex
\subsection{Electronics and data acquisition hardware}

The CUORE electronics will provide an effective low-noise system for reading and monitoring the detectors. The main boards consist of 8-layer $233\times280$~mm$^2$ PCBs which accommodate 6~channels each. Each channel consists of a preamplifier and a programmable-gain amplifier, load resistors, a detector biasing system, as well as a number of other circuits that enable the DC characterization of each thermistor and the monitoring of many voltage nodes. The anti-aliasing Bessel filter boards have programmable cutoff frequencies in the range 15--120~Hz, allowing for optimal analog filtering when used in conjunction with the new DAQ system that has higher sampling rates than the system that was used for Cuoricino and which is currently being used for CUORE-0.
The production, characterization, and calibration of the electronics are in progress.

We are in the process of procuring all CUORE data acquisition~(DAQ) hardware, including National Instruments NI-628x-series high-precision 18-bit digitizers. A small DAQ system, based on a single chassis, will be used for upcoming commissioning tests of the CUORE cryostat while we configure and test the full DAQ.

A Faraday cage will be needed to shield the 
% high-impedance 
signal links between the detectors and the front-end electronics from disturbances coming from the main power line~(50 Hz), the cryogenic pumps, and any other electromagnetic interferences which could be injected from outside. The cage will be located atop the cryostat, on the second floor of the CUORE building, and have a volume of $\sim6\times6\times3\1{m^3}$ and a total surface area of $\sim150\1{m^2}$. 
The design specification is for a 60~dB attenuation at 50~Hz. 
%The mechanical design of the structure is ongoing. 
%The grounding configuration of the whole experiment is also being finalized.

The CUORE slow control system will use LabVIEW for the instrumentation drivers,
% while the TANGO package~\cite{} will be used for high-level systems including user interfaces, clients, web server, and the database. 
while the network layer will use one of the standard protocols available within LabVIEW and will store data in a schema-less database~\cite{books/daglib/0025185} as well as in the CUORE SQL analysis database.
Custom packages will be used for high-level user interfaces, including web-based clients for monitoring and alarms. 
The slow control system is currently under development.

\subsection{Data acquisition and analysis software}

CUORE will use a custom-built DAQ software package named \textsc{Apollo} that has been designed to read signals from $\sim$~1000~bolometers. \textsc{Apollo} will digitize the analog waveforms, run trigger algorithms, and store data for offline analysis. Data are saved in two formats: triggered bolometer pulses are saved to ROOT files~\cite{Brun:1997pa} while the continuous waveforms are saved to compressed ASCII files. Detector parameters and run configurations are stored in an SQL database which is also used for offline data analysis. \textsc{Apollo} provides graphical user interfaces for run control and monitoring, and a slow-control system for interacting with the front-end electronics. It also includes tools for the automated detector characterization to be performed in the start-up phase of the experiment. \textsc{Apollo} has been tested extensively in our experimental setups at Hall A and Hall C, most recently during CUORE-0 data taking.

For data analysis we use a custom-built software framework named \textsc{Diana}, which was developed using a plugin architecture in C{}\verb!++!. \textsc{Diana} has been used extensively as our standard tool for analyzing data from R\&D runs, Cuoricino, and CUORE-0.
As we analyze the CUORE-0 data we are continuing to build upon the standard analysis procedure (see Sec.~\ref{ssec:q0-bg}), developing new features such as noise decorrelation~\cite{ManciniTerracciano:2012fq} and a web-based data quality monitoring system.

% !TEX root = CUORE_status_review_2014.tex
\section{Conclusion and outlook}
\label{sec:conclusion}

\begin{figure}[t]
 \begin{tabular}{lr}
 \subfloat[]{\label{fig:half-life}\includegraphics[height=0.4\columnwidth]{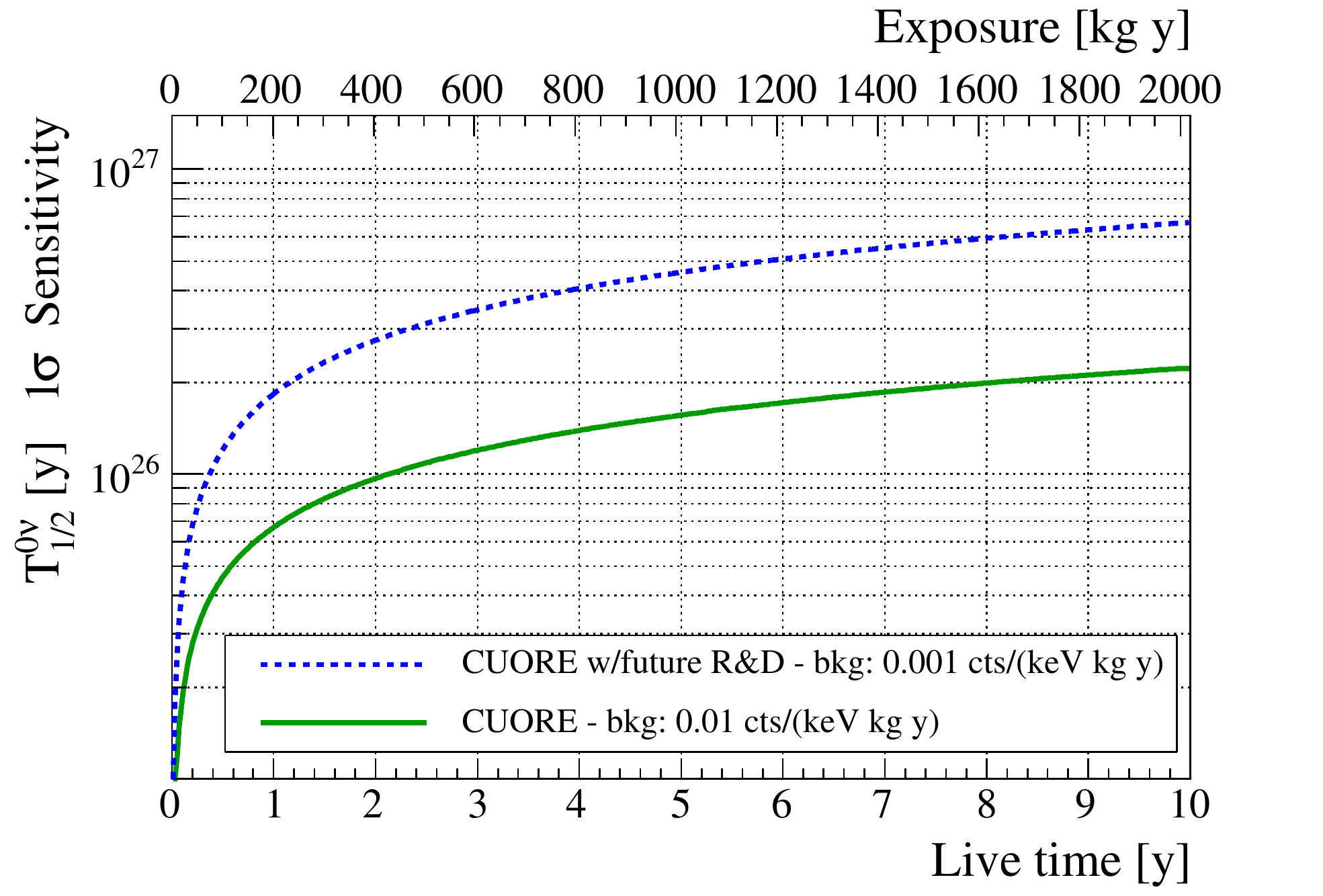}} 
\subfloat[]{\label{fig:mee}\includegraphics[height=0.39\columnwidth]{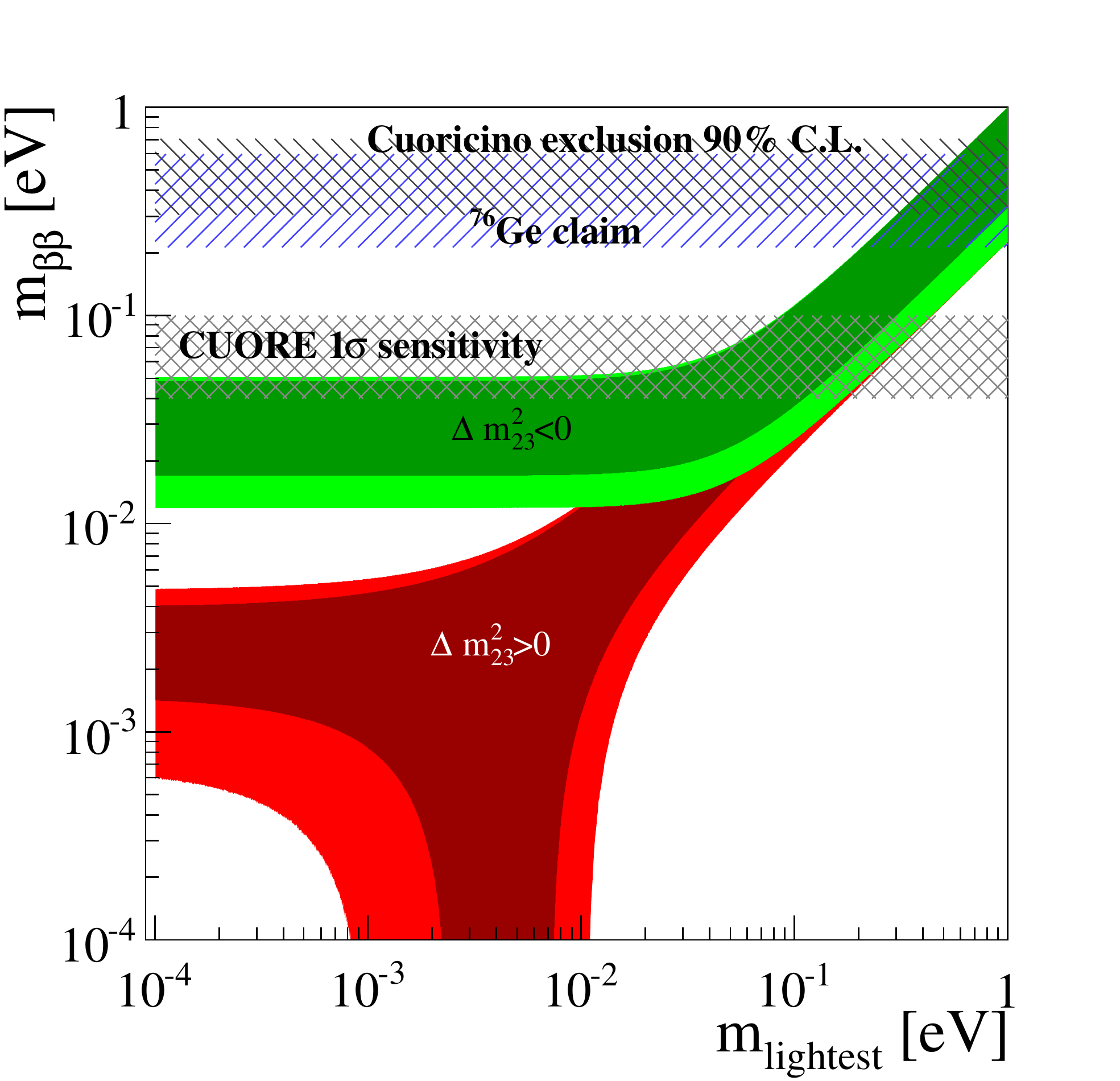}}\\
 \end{tabular}
\caption{(a)~Projected CUORE $1\sigma$ sensitivity to the half-life of $0\nu\beta\beta$~decay of $^{130}$Te as a function of detector live time. The solid green line shows the sensitivity for the target background rate of 0.01~counts/keV/kg/y, while the dashed blue line shows a speculative future scenario in which the background is ten times lower. (b)~The corresponding $1\sigma$ sensitivity to effective Majorana mass versus lightest neutrino mass after five years of detector live time. The spread in the projected CUORE $m_{\beta\beta}$ bands arises from uncertainties in calculations of the nuclear matrix elements used to convert a measured half-life into an effective Majorana mass. The green band labeled $\Delta m^2_{23}<0$ denotes the inverted neutrino mass hierarchy while the red band labeled $\Delta m^2_{23}>0$ indicates the normal mass hierarchy. For both, the darker inner bands represent regions allowed by the best-fit neutrino oscillation parameters in the PMNS matrix, while the lighter outer bands extend to $3\sigma$ coverage. Both figures are from~\cite{CUORE_sensitivity_2011}.}
\end{figure}

With its large detector mass and an excellent anticipated energy resolution, CUORE is one of the most sensitive $0\nu\beta\beta$~decay experiments under construction.
Figure~\ref{fig:half-life} shows the experiment's projected half-life sensitivity to $^{130}$Te $0\nu\beta\beta$~decay as a function of live time, assuming the target background rate of 0.01~counts/keV/kg/y is achieved. After five years of live time CUORE should reach a $1\sigma$ (90\%~C.L.) sensitivity of $1.6\times10^{26}\1{y}\,(9.5\times10^{25}\1{y}$) on the half-life of $0\nu\beta\beta$~decay of $^{130}$Te.

For $0\nu\beta\beta$ decay involving exchange of light Majorana neutrinos, the half life $T^{0\nu}_{1/2}$ can be expressed as
\begin{equation}
(T^{0\nu}_{1/2})^{-1} = G^{0\nu}(Q,Z)|M^{0\nu}|^2\frac{|\langle m_{\beta\beta}\rangle|^2}{m_e^2} ~,
\label{eq:T_mBB}
\end{equation}
where $G^{0\nu}(Q)$ is an accurately calculable phase-space factor which scales with the decay's $Q$-value as $Q^5$; $M^{0\nu}$ is the nuclear matrix element for the process, which carries a large uncertainty due to the range of results calculated from various models; and $|\langle m_{\beta\beta}\rangle|$ is the so-called effective Majorana neutrino mass which correlates $0\nu\beta\beta$~decay with neutrino mixing parameters~\cite{PDG2012}. This formula enables conversion of an experimentally measured $0\nu\beta\beta$~decay half life (or lower limit thereof) into an effective Majorana mass (or upper limit thereof). For CUORE, five years of live time should yield a $1\sigma$ (90\% C.L.) sensitivity to an effective Majorana mass in the range 40--100~meV (50--130~meV), which overlaps the top edge of the allowed band for the inverted mass hierarchy~(Figure~\ref{fig:mee}).
% 5sigma half-life: 3\times10^{25} y, with corresponding effective Majorana mass: 90 - 240 meV

The successful commissioning of CUORE-0 and its promising background rates represent a significant milestone for CUORE. The average energy resolutions of the CUORE-0 bolometers are on par with Cuoricino and are among the best seen in large-mass bolometer arrays, even before optimization and while running at a suboptimal working temperature. We find the background rates in the $\alpha$ continuum region and the ROI are lower by a factor of 6 and 2, respectively, with respect to Cuoricino. This success in reducing the background has confirmed the efficacy of our copper-cleaning techniques and detector-assembly methods. We intend to continue operating CUORE-0 until CUORE begins data taking, by which time CUORE-0 should have become the most sensitive experiment searching for $0\nu\beta\beta$~decay of $^{130}$Te.

CUORE is now in an advanced state of construction, making steady progress in all respects.
Detector assembly was recently completed, and the phased commissioning of the cryostat and integration of its many interconnected systems, 
including the DCS and the dilution unit, is ongoing.
We plan to complete the integration and commissioning of CUORE at the end of 2014 and commence data taking in the first half of 2015.

\section*{Acknowledgments}

\noindent
The CUORE Collaboration thanks the directors and staff of the Laboratori Nazionali del Gran Sasso and the technical staff 
of our laboratories. This work was supported by the Istituto Nazionale di Fisica Nucleare~(INFN); the National Science 
Foundation under Grant Nos.~NSF-PHY-0605119, NSF-PHY-0500337, NSF-PHY-0855314, NSF-PHY-0902171, and 
NSF-PHY-0969852; the Alfred P. Sloan Foundation; the University of Wisconsin Foundation; and Yale University. 
This material is also based upon work supported by the US Department of Energy~(DOE) Office of Science under Contract 
Nos.~DE-AC02-05CH11231 and DE-AC52-07NA27344; and by the DOE Office of Science, Office of Nuclear Physics, 
under Contract Nos.~DE-FG02-08ER41551 and DEFG03-00ER41138. This research used resources of the National 
Energy Research Scientific Computing Center (NERSC).

\bibliographystyle{apsrev} 
\bibliography{cuore_review}
\end{document}